\documentclass[11pt,letterpaper]{article}
\usepackage{caption}
\DeclareCaptionFont{sansfont}{\fontseries{n}\fontfamily{phv}\selectfont}
\makeatletter
\renewenvironment{table}%
  {\renewcommand\familydefault\sfdefault
   \@float{table}}
  {\end@float}
\makeatother

\usepackage[text={180mm, 240mm}]{geometry}

\usepackage{amsmath,amssymb}
\usepackage{changepage, comment, subcaption}
\usepackage[utf8x]{inputenc}
\usepackage{cite}
\usepackage[right]{lineno}
\usepackage{microtype, booktabs}
\usepackage{array}

\usepackage[aboveskip=1pt,labelfont=bf,labelsep=period,justification=raggedright,singlelinecheck=off]{caption}

\makeatletter
\renewcommand{\@biblabel}[1]{\quad#1.}
\makeatother
\usepackage{lastpage,fancyhdr,graphicx}
\usepackage{epstopdf}
\usepackage{pgfplots, tikz}
\pgfplotsset{compat=1.13}
\usepackage{pgfplotstable}
\usetikzlibrary{positioning, calc, arrows, 3d, decorations.pathreplacing}
\usepgfplotslibrary{groupplots}
\pgfplotsset{major grid style={dotted,green!50!black}}

\begin{document}
\vspace*{0.2in}

% Title must be 250 characters or less.
\begin{flushleft}
{\Large
\textbf\newline{A convolutional neural-network model of human cochlear mechanics and filter tuning for real-time applications} % Please use "sentence case" for title and headings (capitalise only the first word in a title (or heading), the first word in a subtitle (or subheading), and any proper nouns).
}
%DEEPAK, we can consider dropping level-dependent in the title if it does not flow enough, even though I think it is important to have level-dependent in the abstract somewhere).
\newline
% Insert author names, affiliations and corresponding author email (do not include titles, positions, or degrees).
\\
Deepak Baby,
Arthur Van Den Broucke,
Sarah Verhulst
\\
\bigskip
Hearing Technology @ WAVES, Dept. of Information Technology, Ghent University, 9000 Ghent, Belgium
\\
\bigskip

% Current address notes
%\textcurrency Current Address: Dept/Program/Center, Institution Name, City, State, Country % change symbol to "\textcurrency a" if more than one current address note
% \textcurrency b Insert second current address 
% \textcurrency c Insert third current address

* s.verhulst@ugent.be; deepakbabycet@gmail.com

\end{flushleft}
% Please keep the abstract below 300 words

\section*{Abstract}
Auditory models are commonly used as feature extractors for automatic speech-recognition systems or as front-ends for robotics, machine-hearing and hearing-aid applications. Although auditory models can capture the biophysical and nonlinear properties of human hearing in great detail, these biophysical models are computationally expensive and cannot be used in real-time applications. We present a hybrid approach where convolutional neural networks are combined with computational neuroscience to yield a real-time end-to-end model for human cochlear mechanics, including level-dependent filter tuning (CoNNear). 
The CoNNear model was trained on acoustic speech material and its performance and applicability were evaluated using (unseen) sound stimuli commonly employed in cochlear mechanics research. The CoNNear model accurately simulates human cochlear frequency selectivity and its dependence on sound intensity, an essential quality for robust speech intelligibility at negative speech-to-background-noise ratios. The CoNNear architecture is based on parallel and differentiable computations and has the power to achieve real-time human performance. These unique CoNNear features will enable the next generation of human-like machine-hearing applications.  

%\linenumbers

\section*{Introduction}

The human cochlea is an active, nonlinear system that transforms sound-induced vibrations of the middle-ear bones to cochlear travelling waves of basilar-membrane (BM) motion \cite{vonbekesy1970}. Cochlear mechanics and travelling waves are responsible for hallmark features of mammalian hearing, including the level-dependent frequency selectivity \cite{narayan1998frequency,robles_2001,shera_2002revised,Oxenham2003} that results from a cascade of cochlear mechanical filters with centre frequencies (CFs) between 20 kHz and 40 Hz from the human cochlear base to apex \cite{Greenwood1990}.

Modelling cochlear mechanics has been an active field of research because computational methods can help characterise the mechanisms underlying normal or impaired hearing and thereby improve hearing diagnostics \cite{jepsen_jasa2011, bondy_sp2004} and treatment \cite{ewert_pma2013,mondol2019}, or inspire machine-hearing applications \cite{lyon_2017, dbaby_interspeech2015}. One popular approach involves representing the cochlea as a transmission line (TL) by discretising the space along the BM and describing each section using a system of ordinary differential equations that approximate the local biophysical filter characteristics (Fig.~\ref{fig:aecnn}a; state-of-the-art model) \cite{de1980auditory, diependaal_jasa1987, zweig_jasa1991, Talmadge1991, Moleti2008, epp_jasa2010, verhulst_jasa2012}. Analytical TL models represent the cochlea as a cascaded system in which the response of each section depends on the responses of all previous sections. This architecture makes them computationally expensive, as the filter operations in the different sections cannot be computed in parallel. The computational complexity is even greater when nonlinearities or feedback pathways are included to faithfully approximate cochlear mechanics \cite{zweig_jasa1991,zweig2016}.

Computational complexity is the main reason that real-time applications for hearing-aid \cite{hohmann_book2008}, robotics \cite{rascon_robotics2017}, and automatic speech recognition applications do not adopt cochlear travelling-wave models in their preprocessing. Instead, they use computationally efficient approximations of auditory filtering that compromise on key auditory features. A common simplification implements the cochlear filters as a parallel, rather than cascaded, filterbank \cite{morgan_book2004, patterson_jasa1995}. However, this parallel architecture captures neither the longitudinal coupling properties of the BM \cite{shera2001frequency} nor the generation of otoacoustic emissions \cite{shera2008mechanisms}. Another popular model, the gammatone filterbank model \cite{hohmann_gammatone2002}, ignores the stimulus-level dependence of cochlear filtering. Lastly, a number of models simulate the level-dependence of cochlear filtering but fail to match the performance of TL models \cite{saremi_jasa2016}: they either simulate the longitudinal cochlear coupling locally within the individual filters of the uncoupled filterbank \cite{DRNL} or introduce distortion artefacts when combining an automatic-gain-control type of level-dependence with cascaded digital filters \cite{CARFAC,saremi_jasa2018, altoe2017dynamics}.   

Thus, the computational complexity of biophysically realistic cochlear models poses a serious impediment to the development of human-like machine-hearing applications. This complexity motivated our search for an efficient model that matches the performance of state-of-the-art analytical TL models while offering real-time execution. Here, we investigate whether convolutional neural networks (CNNs) can be used for this purpose. Neural networks of this type can deliver end-to-end waveform predictions \cite{dbaby_icassp2019, santi_segan_is2017} with real-time properties \cite{fotios_ica2019} and are based on convolutions akin to the filtering process associated with cochlear processing.

This paper details how CNNs can best be connected and trained to approximate the computations performed by TL cochlear models \cite{verhulst_jasa2012, altoe_jasa2014, verhulst_hearres2018}, with a specific emphasis on simultaneously capturing the tuning, level-dependence, and longitudinal coupling characteristics of human cochlear processing. The proposed model (\emph{CoNNear}) converts speech stimuli into corresponding BM displacements across 201 cochlear filters distributed along the length of the BM. Unlike TL models, the CoNNear architecture is based on parallel CPU computations that can be accelerated through GPU computing. Consequently, CoNNear can easily be integrated with real-time auditory applications that use deep learning. The quality of the CoNNear predictions and the generalisability of the method are evaluated on the basis of cochlear mechanical properties such as filter tuning estimates \cite{Oxenham2010}, nonlinear distortion characteristics \cite{robles1991two}, and spatial excitation patterns \cite{ren2002longitudinal} obtained using sound stimuli commonly adopted in experimental studies of cochlear mechanics but not included in the training. 

\section*{The CoNNear model}

\begin{figure*}[t!]
\begin{center}
\includegraphics[scale=1]{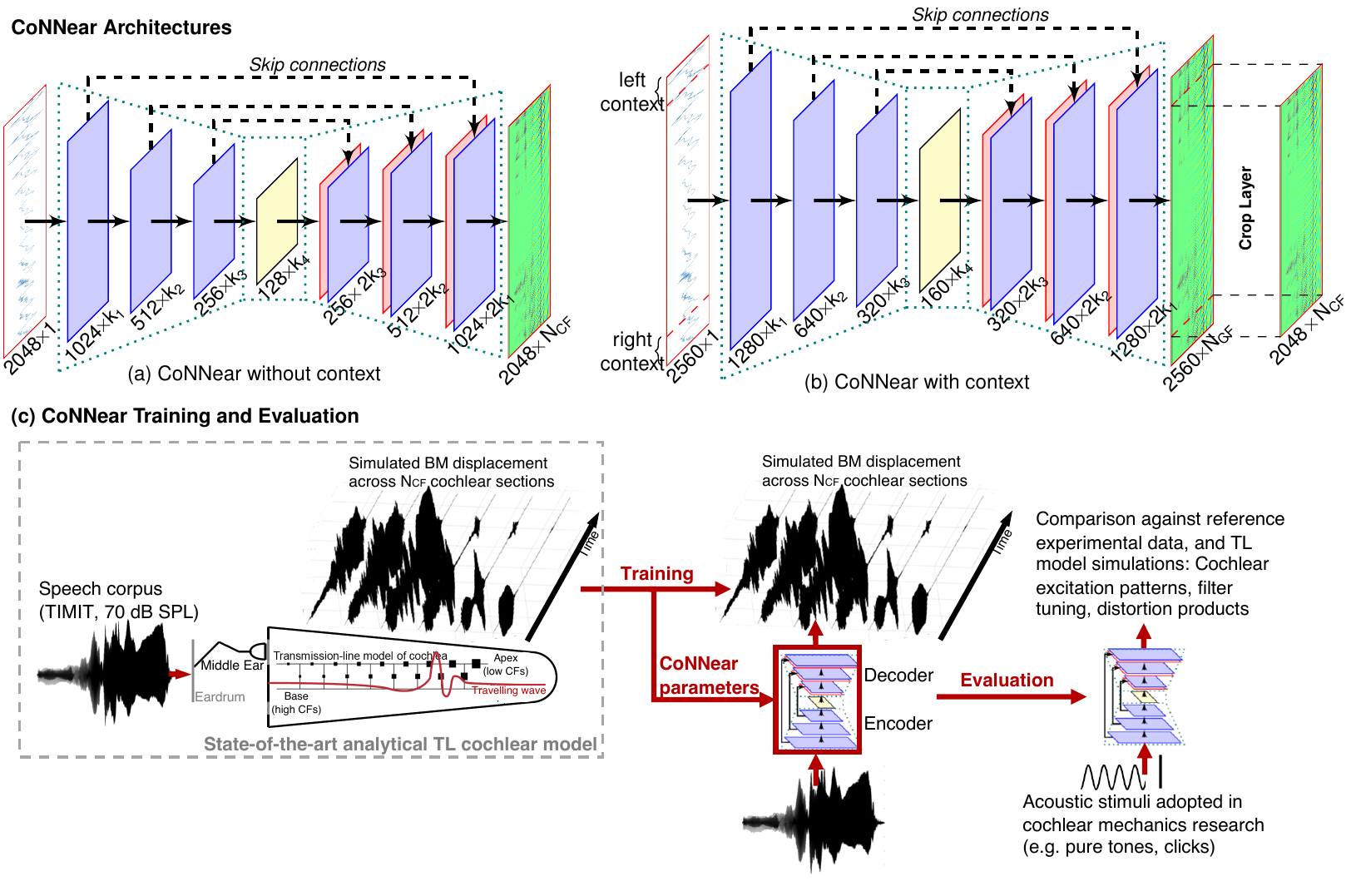}
\end{center}
\vspace{-10pt}
\caption{\textbf{CoNNear Overview.} CoNNear is a fully convolutional encoder-decoder neural network with strided convolutions and skip-connections to map audio input to 201 basilar-membrane vibration outputs of different cochlear sections (N$_{\text{CF}}$) in the time-domain. CoNNear architectures with (a) and without (b) context are shown. The final CoNNear model has four encoder and decoder layers, uses context and includes a tanh activation function between the CNN layers. (c) provides an overview of the model training and evaluation procedure. Whereas reference, analytical TL-model simulations to a speech corpus were used to train the CoNNear parameters, evaluation of the model was performed using simple acoustic stimuli commonly adopted in cochlear mechanics studies.}
\label{fig:aecnn}
\end{figure*}

The CoNNear model has an auto-encoder CNN architecture and transforms a 20-kHz sampled acoustic waveform (in [Pa]) to N\textsubscript{CF} cochlear BM displacement waveforms (in [$\mu$m]) using several CNN layers and dimension changes (Fig.~\ref{fig:aecnn}a). The first four layers are \emph{encoder} layers and use strided convolutions to halve the temporal dimension after every CNN layer. The following four, \emph{decoder}, layers map the condensed representation onto $L \times N_{\text{CF}}$ outputs using deconvolution operations. $L$ corresponds to the initial size of the audio input and $N_{\text{CF}}$ to 201 cochlear filters with centre frequencies (CFs) between 0.1 and 12 kHz. The adopted CFs were spaced according to the Greenwood place-frequency map of the cochlea \cite{Greenwood1990} and span the most sensitive frequency range of human hearing \cite{ISO2003}. It is important to preserve the temporal alignment (or, phase) of the inputs across the architecture, because this information is essential for speech perception \cite{lorenzi2006speech}. We used \emph{U}-shaped skip connections for this purpose. Skip connections have earlier been adopted in image-to-image translation \cite{pix2pix} and speech-enhancement applications \cite{dbaby_icassp2019, santi_segan_is2017}; they pass temporal information directly from encoder to decoder layers (Fig.~\ref{fig:aecnn}a; dashed arrows). Aside from preserving phase information, skip connections may also improve the model's ability to learn how best to combine the nonlinearities of several CNN layers to simulate the level-dependent properties of human cochlear processing.

Every CNN layer is comprised of a set of filterbanks followed by a nonlinear operation \cite{lecun_cnn2015} and the CNN filter weights were trained using TL-simulated BM displacements from N\textsubscript{CF} cochlear channels \cite{verhulst_hearres2018}. While training was conducted using a speech corpus \cite{timit} presented at 70 dB SPL, model evaluation was based on the ability to reproduce key cochlear mechanical properties using basic acoustic stimuli (e.g. clicks, pure-tones) unseen during training (Fig.~\ref{fig:aecnn}c). During training and evaluation, the audio input was segmented into $2048$-sample windows ($\approx$100 ms), after which the corresponding BM displacements were simulated and concatenated over time. Because CoNNear treats each input independently, and resets its adaptation properties at the start of each simulation, this concatenation procedure could result in discontinuities near the window boundaries. To address this issue, we also evaluated an architecture that had the previous and following ($256$) input samples available as context (Fig.~\ref{fig:aecnn}b). Different from the no-context architecture (Fig.~\ref{fig:aecnn}a), a final cropping layer was added to remove the simulated context and yield the final $L$-sized BM displacement waveforms. Additional details on the CoNNear architecture and training procedure are given in Methods and Extended Data Fig.~1. Lastly, training CoNNear using audio inputs of fixed duration does not prevent it from handling inputs of other durations after training, thanks to its convolutional architecture. This flexibility is a clear benefit over matrix-multiplication-based neural network architectures, which can operate only on inputs of fixed-duration.

\section*{CoNNear hyperparameter tuning}

Critical to this work is proper determination of the optimal CNN architecture and its hyperparameters that yield a realistic model for cochlear processing. Ideally, CoNNear should both simulate the speech training dataset with a sufficiently low L1 loss [i.e., the average mean-absolute difference between simulated BM displacements of the reference TL and CoNNear model] and also reproduce key cochlear response properties. Details on the hyperparameters that can be adjusted to define the final CoNNear architecture (number of layers, nonlinearity, context) are given in Extended Data Fig.~1. To determine the hyperparameters, we followed an iterative principled approach taking into account: (i) the L1 loss on speech material, (ii) the desired frequency- and level-dependent cochlear filter tuning characteristics, and (iii), the computational load needed for real-time execution. Afterwards, the prediction accuracy of CoNNear was evaluated on a broader set of response features. We considered a total of four metrics derived from cochlear responses evoked using simple stimuli of different frequencies and levels: filter tuning (Q\textsubscript{ERB}), click-evoked dispersion, pure-tone excitation patterns, and the generation of cochlear distortion products. Together, these evaluation metrics form a comprehensive description of cochlear processing. Further information on the tests and their implementation is given in Methods. A benefit of our approach is that the evaluation stimuli were \emph{unseen} to the model to ensure an independent evaluation procedure. Even though any fragment of the speech training material can be seen as a combination of basic acoustic elements such as impulses and pure tones of varying levels and frequencies, the cochlear mechanics stimuli were not explicitly present in the training material.

\begin{figure*}[h!]
\begin{center}
\includegraphics[scale=1.0]{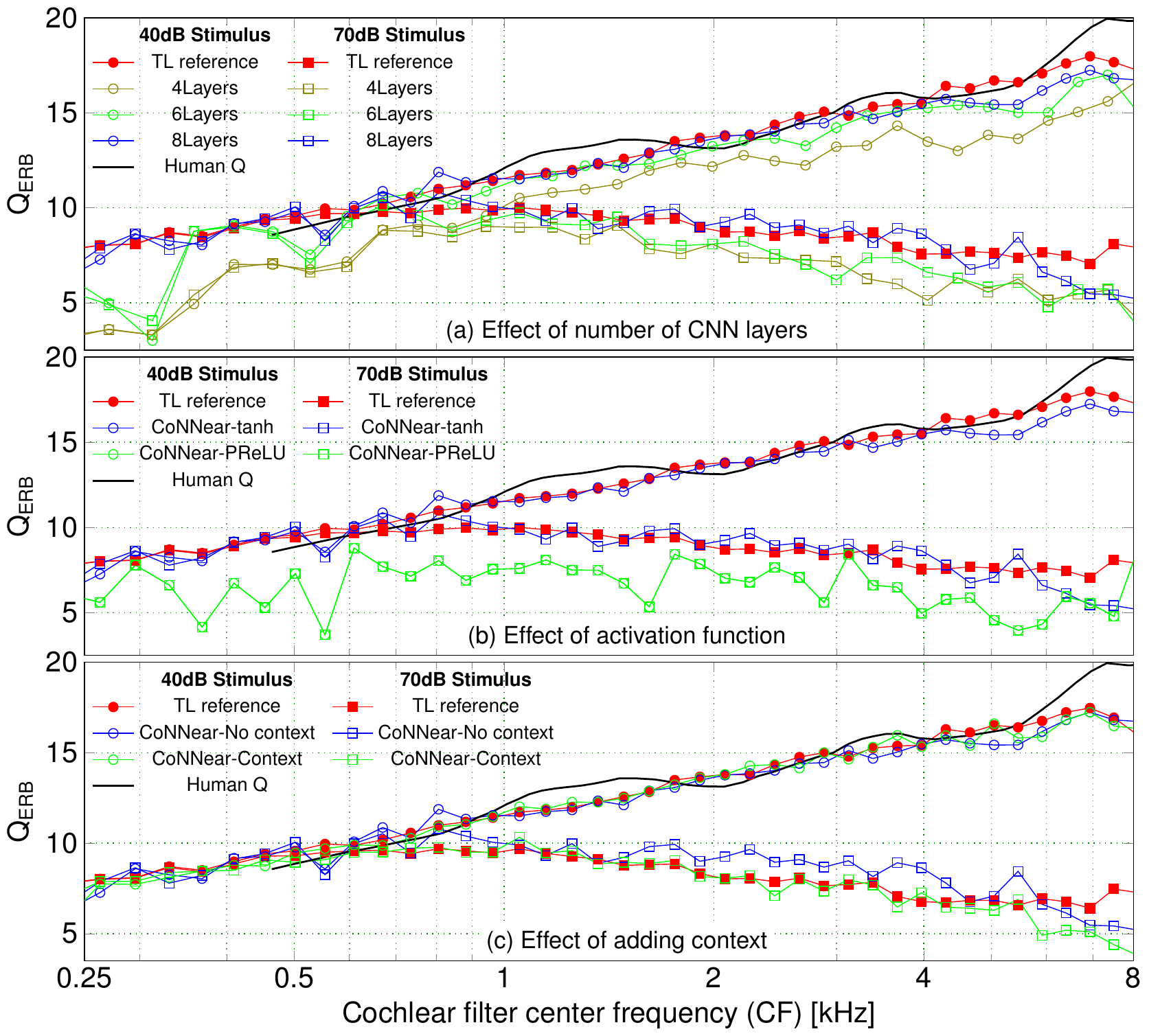}
\end{center}
\caption{\textbf{CoNNear hyperparameter tuning.} Cochlear filter tuning (Q\textsubscript{ERB}) was simulated to 100-$\mu$s clicks of $40$ and $70$ dB peSPL for different cochlear filter centre frequencies (CF). Reference human Q\textsubscript{ERB} estimates for low-stimulus levels \cite{shera2010otoacoustic} are shown and compared to simulations of the reference TL model and CoNNear models with (c) and without (a,b) context. (a) Increasing the number of CoNNear layers from $4$ to $8$, improved the Q\textsubscript{ERB}-across-CF simulations. (b) Comparing architectures with the PReLU or tanh activation function shows that the PReLU nonlinearity failed to capture level-dependent cochlear filter tuning because the Q\textsubscript{ERB} functions remained invariant to stimulus level changes. $8$-layer CNN architectures were used for this simulation. (c) Adding context to the $8$-layer, tanh model further improved the CoNNear predictions by showing less Q\textsubscript{ERB} fluctuations across CF.}
\label{fig:hyperparameters}
\end{figure*}

The first hyperparameter specifies the total number of encoder/decoder layers, which we set to $8$ on the basis of Q\textsubscript{ERB} simulations. Aside from the reference experimental human Q\textsubscript{ERB} curve for low stimulus levels \cite{shera2010otoacoustic}, Fig.~\ref{fig:hyperparameters}a shows simulated Q\textsubscript{ERB} curves from the reference TL-model (red) overlaid with CoNNear-model simulations. CoNNear captures the frequency-dependence of the human and reference TL-model Q\textsubscript{ERB} function better as the layer depth was increased from $4$ to $8$. Models with $4$ and $6$ layers underestimated the overall Q\textsubscript{ERB} function and performed worse for CFs below 1 kHz where reference ERBs were narrower, and corresponding BM impulse responses longer, than for higher CFs. Extending the number of layers beyond $8$ increased the required computational resources without producing a substantial improvement in the quality of the fits. These resources are listed in Extended Data Fig.~2 for models of different layers, along with L1-loss predictions on the speech training set and a small set of evaluation metrics.  

\begin{figure*}[h!]
{\begin{center}
\includegraphics[scale=0.9]{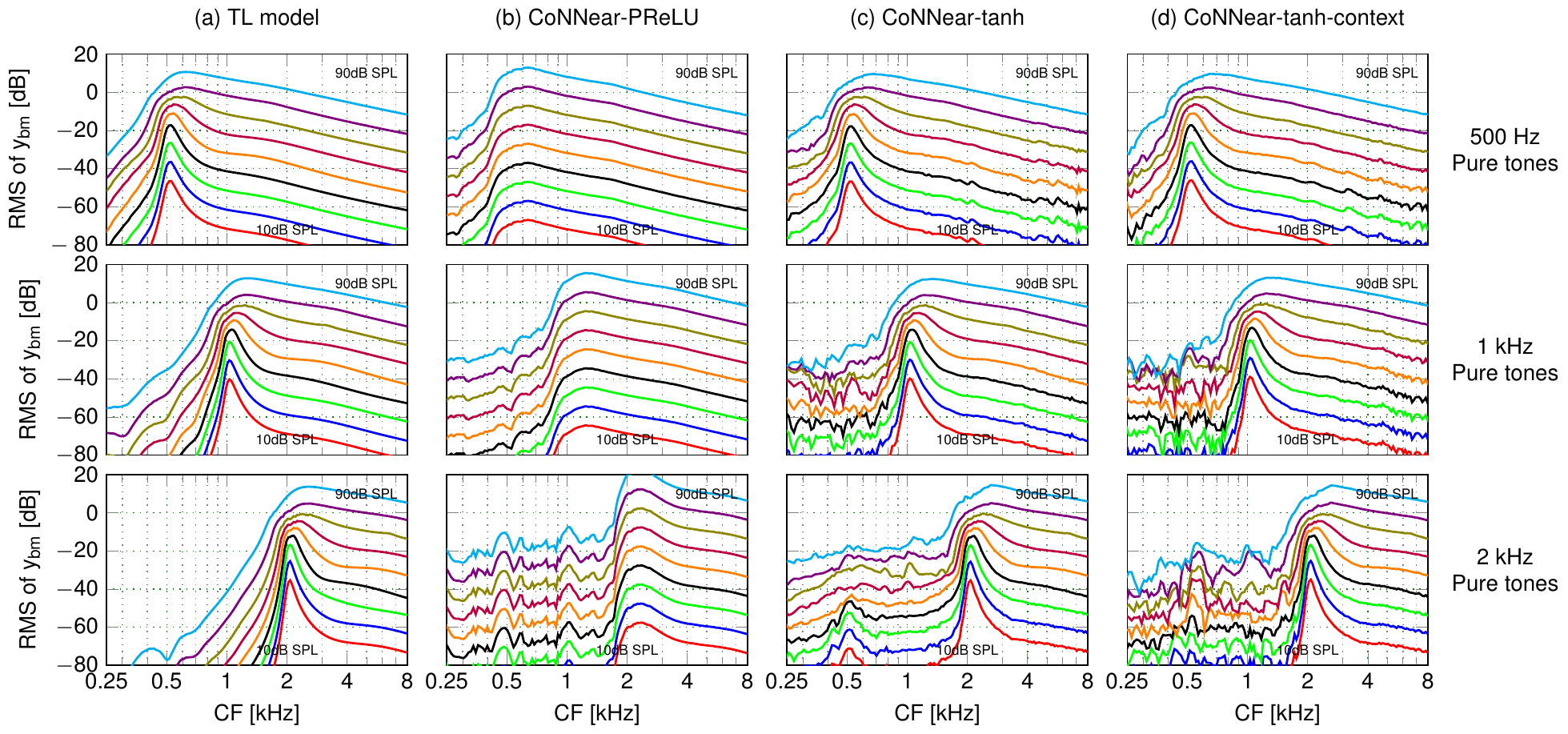}
\end{center}
}
\caption{\textbf{Comparing cochlear excitation patterns across model architectures.} Simulated root-mean-square (RMS) levels of BM displacement across CF for pure-tone stimuli between 10 and 90 dB SPL. From top to bottom, the stimulus frequencies were $500$ Hz, $1$ kHz and $2$ kHz, respectively. Both the reference, TL-model, level-dependent excitation pattern shape changes and compressive growth of pattern maxima (a), were captured by the tanh architectures (c,d), but not by the PReLU architecture (b).}
\label{fig:expat}
\end{figure*}

The second hyperparameter controls the activation function, or nonlinearity, which is placed between the CNN layers. To mimic the original shape of the outer-hair-cell input/output function \cite{russell1986responses} responsible for cochlear compression, we required that the activation function cross the x-axis. We therefore considered only the parametric rectified linear unit (PReLU) and hyperbolic-tangent (tanh) nonlinearities rather than standard activation functions such as the sigmoid and ReLU. Figure~\ref{fig:hyperparameters}b depicts how the PreLU and tanh activation functions affected the simulated Q\textsubscript{ERB}s across CFs. Whereas the PReLU activation function was unable to capture the level-dependence of cochlear filter tuning, the tanh nonlinearity reproduced both the level and frequency-dependence of human Q\textsubscript{ERB}s. Additionally, Extended Data Fig.~3 (no context simulations) shows that the model with tanh nonlinearity reached a lower L1 loss on both training and test sets compared to the PReLU nonlinearity, while requiring a smaller number of parameters. The benefit of the tanh nonlinearity is further illustrated when comparing simulated pure-tone excitation patterns across model architectures. Figure~\ref{fig:expat} shows that the tanh nonlinearity (c) outperforms the PReLU nonlinearity (b) in capturing the compressive growth of BM displacement with stimulus level, as observed in the excitation pattern maxima (a). Although both activation functions were able to code negative input deflections, only the model with tanh nonlinearity (whose shape best resembles the cochlear input/output function) performed well. These simulations show that it is essential to consider the shape of the activation function when the reference system is nonlinear.

\begin{figure*}[h!]
\centering
\includegraphics[scale=1]{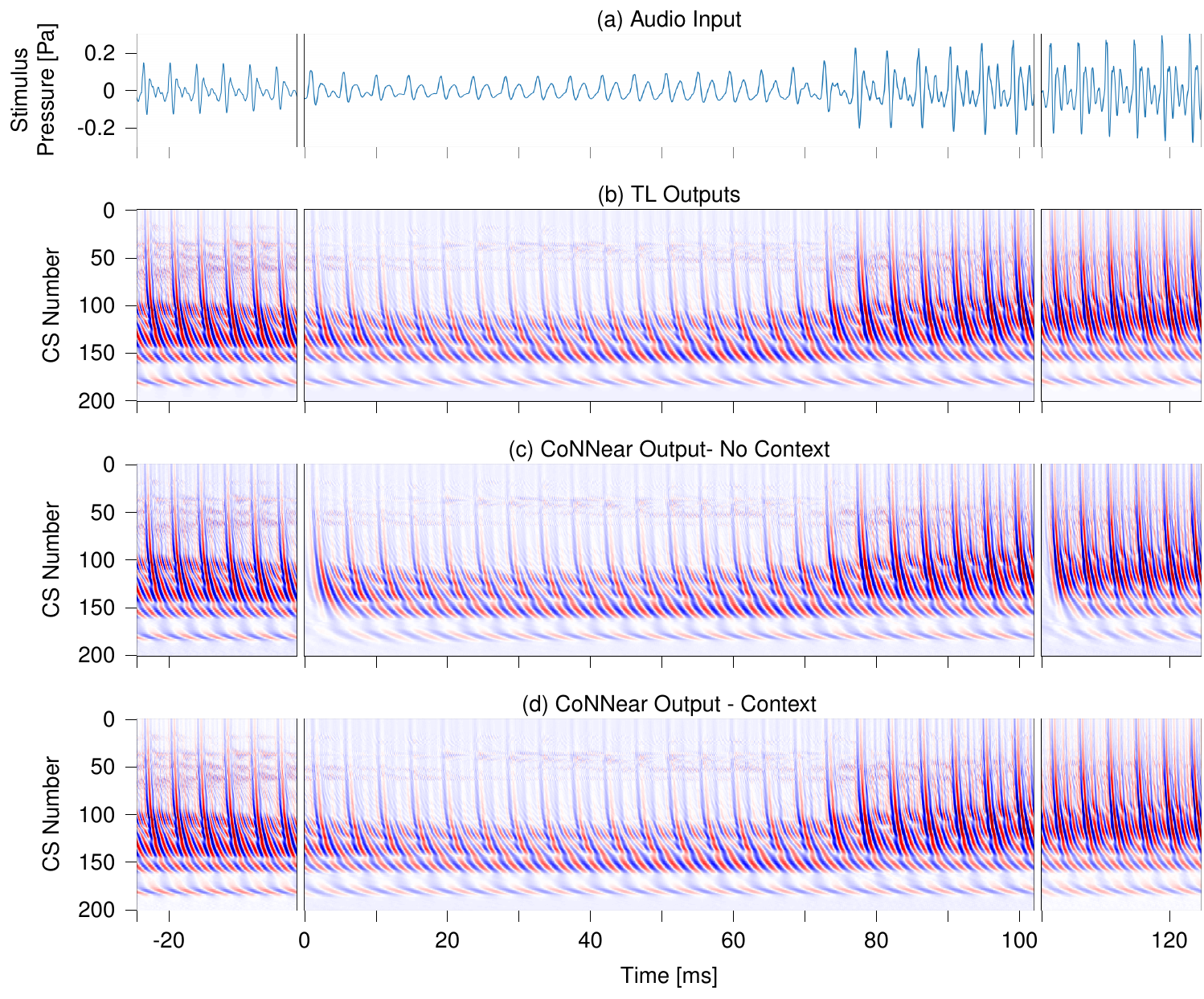}
%\vspace{-10pt}
\caption{\textbf{Effect of adding context to the CoNNear simulations}. Simulated BM displacements for a 2048-sample-long speech fragment of the TIMIT test set (i.e., unseen during training). The stimulus waveform is shown in panel (a) and panels (b)-(d) depict instantaneous BM displacement intensities (darker colours = higher intensities) of simulated TL-model outputs (b) and two CoNNear architecture outputs: without (c) and with (d) context. The left and right segments show the output of the neighbouring windows to demonstrate the discontinuity over the boundaries. The N\textsubscript{CF}=201 output channels are labelled per channel number: channel 1 corresponds to a CF of 12 kHz and channel 201 to a CF of 100 Hz. Intensities varied between -0.5 $\mu$m (blue) and 0.5 $\mu$m (red) for all panels.}
\label{fig:speechinput}
\end{figure*}

The final hyperparameter relates to adding context information when simulating an input of size $L$. As noted earlier, CNN architectures treat each input independently, which can produce discontinuities near the window boundaries. The effect of adding context [cf. architecture (b) vs (a) in Fig.~\ref{fig:aecnn}] is best observed when simulating speech. Figure~\ref{fig:speechinput} shows simulated BM displacements of the reference and CoNNear models to a segment of the speech test set which was not seen during training. Panel (d) shows that providing context prevents discontinuities near the window boundaries. Context was also beneficial when simulating the reference 70-dB Q$_{\text{ERB}}$ function in Fig.~\ref{fig:hyperparameters}c. The L1-loss simulations in Extended Data Fig.~3 (tanh simulations) confirmed the overall performance improvement when providing context, and hence the final CoNNear architecture included a context window of $256$ samples on either side of the $2048$ sample input (Fig.~\ref{fig:aecnn}b).   

\section*{Generalisability of CoNNear}
Overfitting occurs when the trained model merely memorises the training material and fails to generalise to data not present in the training set. To investigate whether the final CoNNear architecture was robust against overfitting, we tested how well the trained CoNNear model performed on unseen stimuli from (i) the same database, (ii) a different database of the same language, (iii) a different database of a different language, and (iv) a music piece.

\begin{figure*}[ht!]
{\begin{center}
\includegraphics[scale=0.6]{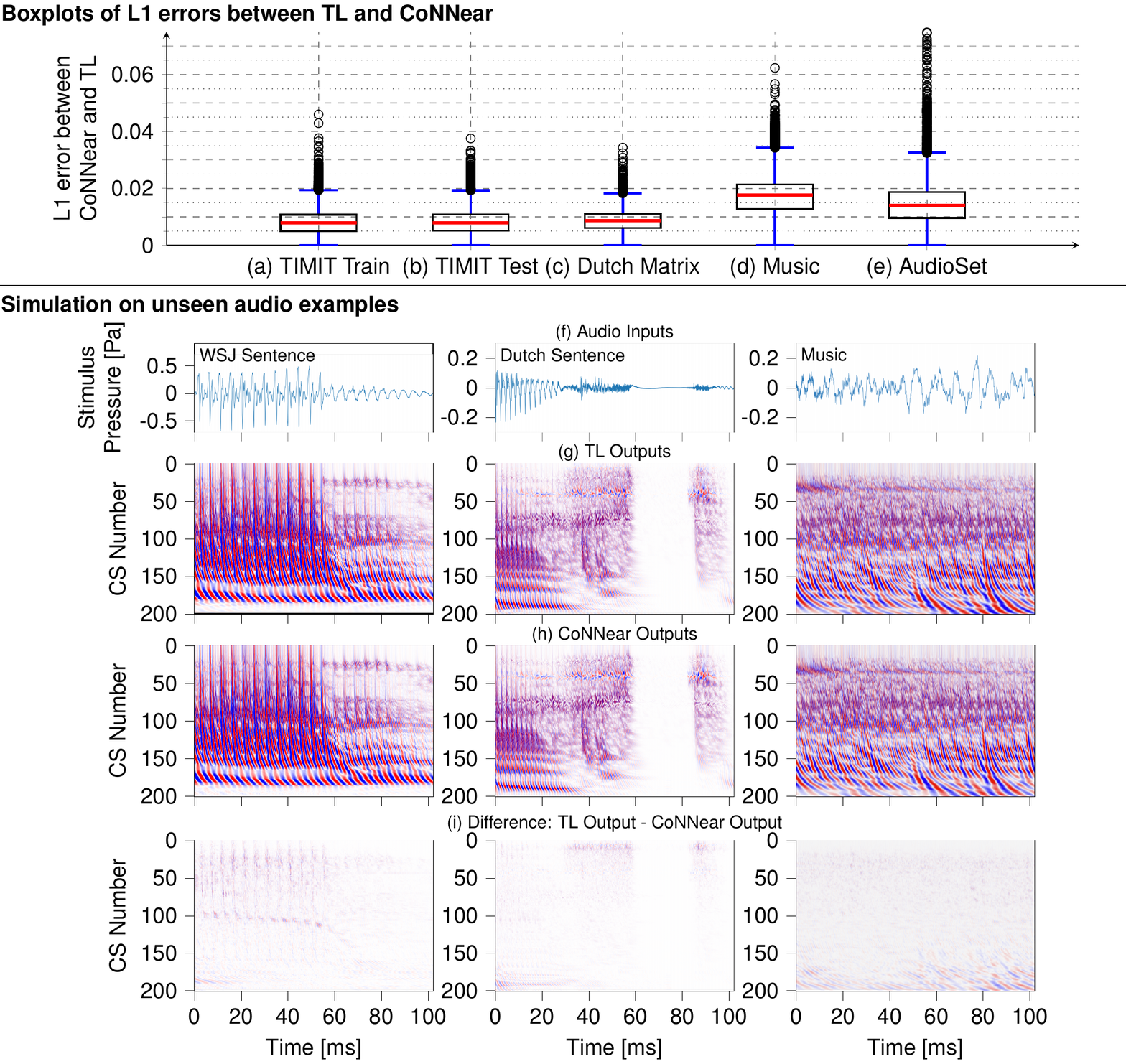}
\end{center}
}
\vspace{-50pt}
\caption{\textbf{Generalisability of CoNNear to unseen input. Top:} boxplots comparing the distribution of L1 losses between TL-model and CoNNear model simulations for $2048$-long windows within (a) 2310 sentences from the TIMIT training set (131760 calculated L1 losses, 16-kHz sampled audio) (b) the 550 sentences in the TIMIT test set (32834 L1 losses, 16-kHz sampled audio), (c) 100 sentences from the Dutch Matrix test (7903 L1 losses, 44.1-kHz sampled audio), (d) All songs of Radioheads OK Computer (48022 L1 losses, 16-kHz downsampled audio), (e) 400 random audio-fragments from the natural-sounds AudioSet database (53655 L1 losses, 16-kHz downsampled audio). The red line in the boxplot shows the median of the L1 errors, the black box denotes the values between the 25\% (Q1) and 75\% (Q3) quartiles. The blue whiskers denote the error values that fall within the 1.5 * (Q3 - Q1) range. Outliers are shown as circles. Cochlear model predictions and prediction errors for the windows associated with median and maximum L1 loss in the TIMIT test set are shown in Extended Data Fig.~5. \textbf{Bottom:} Along the columns, simulated BM displacements for three different $2048$-long audio stimuli are shown: a recording of the English Wall Street Journal speech corpus \cite{wsj}, a sentence from the Dutch matrix test \cite{houben2014development} and a music fragment taken from Radiohead - No surprises. Stimulus waveforms are depicted in panel (f) and instantaneous BM displacement intensities (darker colours = higher intensities) of the simulated TL-model and CoNNear outputs are depicted in (g-h). Panel (i) shows the intensity difference between the TL and CoNNear outputs. The N\textsubscript{CF}=201 considered output channels are labelled per channel number: channel 1 corresponds to a CF of 12 kHz and channel 201 to a CF of 100 Hz. The same colour map was used for all figures and ranged between -0.5 $\mu$m (blue) and 0.5 $\mu$m (red).}
\label{fig:histogram_and_examples}
\end{figure*}

Figure \ref{fig:histogram_and_examples} shows boxplots of the L1-loss distributions of all simulated windows in the following audio material (from left to right): the TIMIT training set, the TIMIT test set using different speakers, the Dutch Matrix sentence database \cite{houben2014development}, Radioheads' OK Computer album, and 400 randomly-drawn samples from different event categories in the AudioSet database \cite{gemmeke2017audio}. In addition to calculated loss distributions, Fig.~\ref{fig:histogram_and_examples} compares instantaneous BM displacement intensities between the reference TL-model (g) and CoNNear (h) for three stimuli unseen during training: an English sentence from the Wall Street Journal Corpus \cite{wsj} to show how CoNNear adapts to different recording settings (first column). A sentence from the Dutch matrix test \cite{houben2014development} to investigate performance on a different language stimulus (middle column), and a music segment to test performance on a non-speech acoustic stimulus (right column). From an application perspective, we also tested how CoNNear handles audio input of arbitrary length. Extended Data Fig.~4 shows that CoNNear generalises well to an unseen speech stimulus of length $0.5$ s ($10048$ samples) and a music stimulus of length $0.8$ s ($16384$ samples), even though training was performed using shorter $2048$-sample windows of fixed duration. 

The similar loss distributions across tested speech conditions, together with the small intensity prediction errors seen in Fig.~\ref{fig:histogram_and_examples}(i), show that CoNNear generalises well to unfamiliar stimuli. When comparing L1-loss distributions between speech (b,c) and non-speech (d,e) audio, we observe increased mean L1-losses along with a greater performance variability, which reflects the large variety of spectral and temporal features present in music or the AudioSet database. Even though median performance is still acceptable for non-speech audio, we conclude that CoNNear performs most comparable to the original TL-model simulations for speech. CoNNear's prediction accuracy on non-speech audio may be improved by retraining with a larger and more-diverse stimulus set than the adopted TIMIT training set. However, a clear benefit from our hybrid, neuroscience-inspired training method is that we can yield a generalisable CoNNear architecture even when using a training dataset which is much smaller than those adopted in standard data-driven NN-approaches \cite{lecun_cnn2015}. Lastly, we note that the non-speech audio were downsampled to 16 kHz to match the frequency content of the TIMIT speech material to allow for a fair comparison between speech and non-speech audio. We can hence not guarantee that CoNNear performs well on non-speech audio with frequency content above 8 kHz.

\section*{CoNNear as a model for human cochlear signal processing}

Since the primary goal of this work was to develop a neural-network based model for human cochlear mechanics, we evaluated how well the trained CoNNear model collectively performs on simulating key cochlear mechanics metrics (see Methods for a detailed description). Figure~\ref{fig:hyperparameters}c shows that the final CoNNear architecture with context simulates the frequency- and level-dependence of human cochlear tuning. At the same time, CoNNear faithfully captures the shape and compression properties of pure-tone cochlear excitation patterns (Fig.~\ref{fig:expat}d). However, we did observe small excitation pattern fluctuations at CFs below the stimulus frequency for the 1 and 2 kHz simulations (middle and bottom row in Fig.~\ref{fig:expat}). These fluctuations had levels of $\approx$30 dB below the peak of the excitation pattern and are hence not expected to degrade the sound-driven CoNNear response to complex stimuli such as speech. This latter statement is supported by (i) the CoNNear speech simulations in Fig.~\ref{fig:histogram_and_examples}i, which show minimally visible noise in the error patterns, and (ii) the root mean-square error (RMSE) observed between simulated TL and CoNNear model excitation patterns for a broad range of frequencies and levels (Extended Data Fig.~6c). To obtain a meaningful error estimate, we normalised the RMSE to the TL excitation pattern maximum to yield a CoNNear error percentage. The error for the final CoNNear architecture remained below 5 \% for pure-tone frequencies up to 4 kHz, and stimulation levels below 80 dB. Increased errors were observed for higher stimulus levels and at CFs of 8 and 10 kHz and stemmed from the overall level and frequency content of the speech material used for training; the dominant energy in speech occurs below 5 kHz, and we presented the speech corpus at 70 dB. However, it is noteworthy that the visual difference between simulated high-level excitation patterns was mostly associated with low-level fluctuations at CFs below the stimulation frequency and not at the stimulation frequency itself (cf. the first and last columns of Fig.~\ref{fig:expat}). This means that the stimulus-driven error that occurs when using broadband stimuli is likely smaller than that observed here for tonal stimuli. In summary, the error appears acceptable for processing speech-like audio. We speculate that CoNNear's application range can be extended by retraining with stimuli at different levels and/or with greater high-frequency content.

\begin{figure}[h!]
{\centering
\includegraphics[scale=0.8]{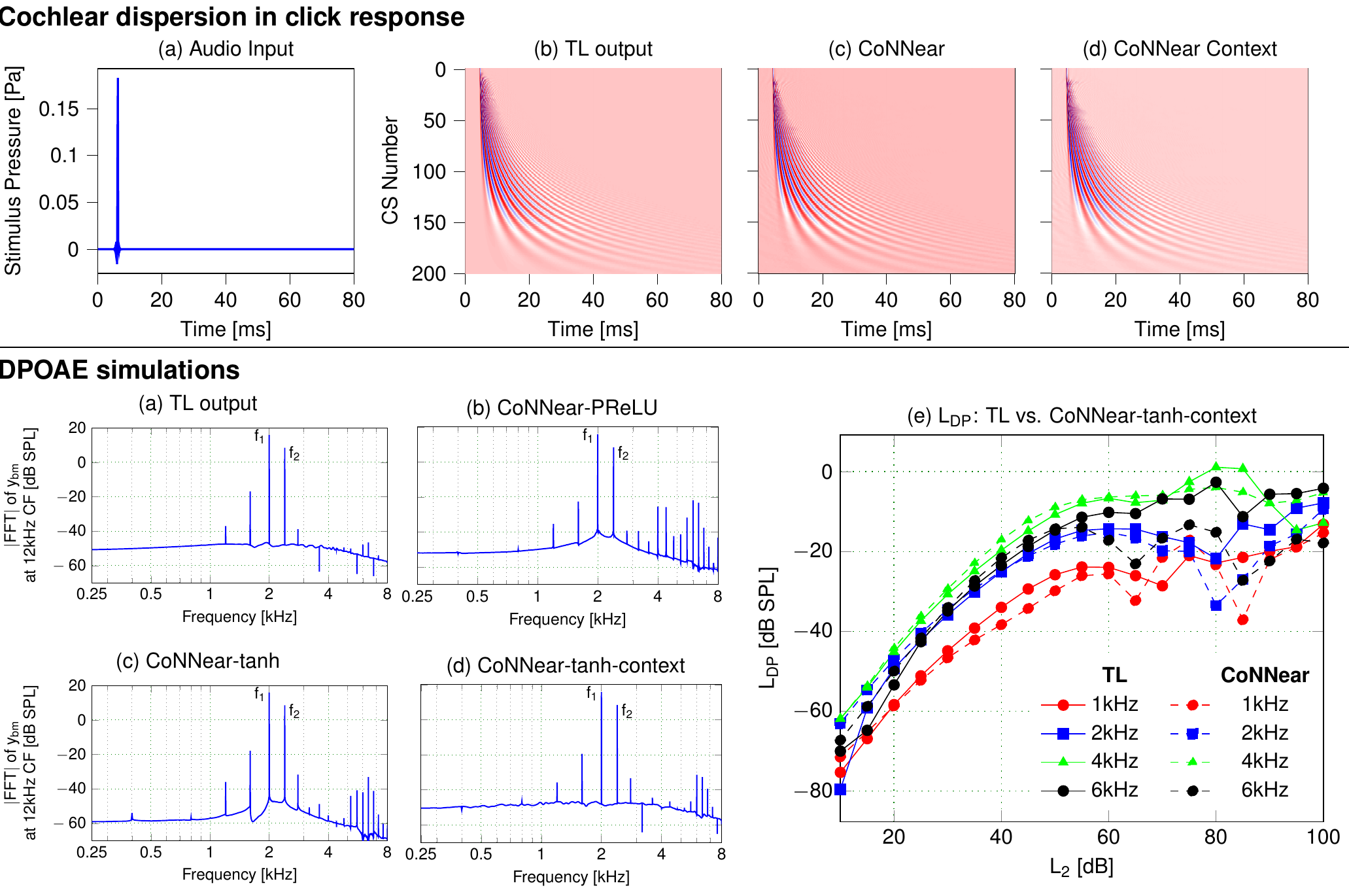}
}
\caption{\textbf{Cochlear dispersion and DPOAEs. Top:} Comparing cochlear dispersion properties across model architectures. Panel (a) shows the stimulus pressure, while panels (b)-(d) show instantaneous BM displacement intensities for CFs (channel numbers, CS) between 100 Hz (channel 201) and 12 kHz (channel 1). The colour scale is the same in all figure panels, and ranges between -15 $\mu$m (blue) and 15 $\mu$m (red). \textbf{Bottom:} Comparing simulated DPOAEs across model architectures. The frequency response of the 12-kHz CF channel (i.e., fast Fourier transform of the BM displacement waveform) was considered as a proxy for otoacoustic emissions recorded in the ear-canal. Panels (a) - (d) show simulations in response to two pure tones of $f_{1}$ of 2.0 and $f_{2}$ of 2.4 kHz for different model architectures. In humans, the most pronounced distortion product occurs at $2f_{1}-f_{2}$ (1.6 kHz). Panel (e) depicts simulated distortion-product levels (L$_{DP}$) compared between TL and the CoNNear-tanh-context model. L$_{DP}$ was extracted from the frequency response of the 12-kHz CF channel at $2f_{1}-f_{2}$. Simulations were conducted for L$_{2}$ levels between 10 and 100 dB SPL and L$_1$=0.4 L$_2$+39 \cite{kummer_scissors}. f$_{1}$ ranged between 1 and 6 kHz, following a f$_{2}$/f$_{1}$ ratio of 1.2.}
\label{fig:clickinput_dpoae}
\end{figure}

The top panel of Fig.~\ref{fig:clickinput_dpoae} shows the dispersion characteristics of the TL model and trained CoNNear models, illustrating the characteristic 12-ms onset delay from basal (high CF, low channel numbers) to apical (low CF, high channel numbers) cochlear sections in both models. In the cochlea, dispersion arises from the biophysical properties of the coupled BM (i.e., its spatial variation of stiffness and damping), and the CoNNear architecture captures this phenomenon. Adding context did not affect the simulation quality.

Finally, we tested CoNNear's ability to simulate a last, key feature of cochlear mechanics: distortion products, which travel in reverse along the BM to generate pressure waveforms in the ear canal (i.e., distortion-product otoacoustic emissions, DPOAEs). The bottom panel of Fig.~\ref{fig:clickinput_dpoae} compares reference TL-model simulations with those of different CoNNear models (b)-(d). DPOAE frequencies are visible as spectral components that are not present in the stimulus (which consists of two pure tones at frequencies $f_1 = 2.0$ kHz and $f_2 = 2.4$ kHz). In humans, the strongest DP component occurs at $2f_1 - f_2 = 1.6$ kHz, and as the simulations show, the level of this DP was best captured using the tanh activation function. As observed before in the excitation pattern simulations, the activation function most resembling the shape of the cochlear nonlinearity performed best when simulating responses that rely on cochlear compression. Adding context removed the high-frequency distortions that were visible in panels (b) and (c).

The quality of the DPOAE simulations across a range of stimulation levels and frequencies was further investigated in Fig.~\ref{fig:clickinput_dpoae}e. Up to L$_2$ levels of 60 dB SPL, CoNNear matched the characteristic nonlinear growth of both TL-simulated and human DPOAE level functions well \cite{dorn2001distortion}. At higher stimulation levels, DPOAE levels started fluctuating in both models. Although this may reflect the chaotic character of cochlear DPs interacting with cochlear irregularities in TL models and in human hearing \cite{shera2008mechanisms}, it may simply reflect the limited training material we had available for CoNNear at these higher levels. Given that human DPOAEs are generally recorded for stimulation levels below 60 dB SPL in clinical or research contexts \cite{janssen2008otoacoustic}, we can conclude that CoNNear was able to capture this important epiphenomenon of hearing. 
The cochlear mechanics evaluations we performed in Figs.~\ref{fig:hyperparameters}, \ref{fig:expat}, \ref{fig:clickinput_dpoae} and Extended Data Fig.~6 together demonstrate that the 8-layer, tanh, CoNNear model with context performed best on four crucial aspects of human cochlear mechanics. Despite training on a limited speech corpus presented at 70 dB SPL, CoNNear learned to simulate outputs which matched those of biophysically-realistic analytical models of human cochlear processing across level and frequency.    

\section*{CoNNear as a real-time model for audio applications}
In addition to its ability to simulate realistic cochlear mechanical responses, CoNNear operates in real-time. In audio applications, real-time is commonly defined as a computation duration less than 10 ms; below this limit no delay is perceived. Table~\ref{tab:timing} summarises the necessary time to compute the final CoNNear-context model for a stimulus window of 1048 samples on CPU or GPU architectures. On a CPU, the CoNNear model outperforms the TL-model by a factor of 129 and on a GPU, CoNNear is 2142 times faster. Additionally, the GPU computations show that the trained CoNNear model (8-layers, tanh, with context) has a latency of 7.27 ms, and hence reaches real-time audio processing performance. 

\begin{table}[h!]
\begin{center}
\caption[Timing of models]{\textbf{Model calculation speed.} Comparison of the time required to calculate a TL and CoNNear model window of 1048 samples on a CPU (Apple MacBook Air, 1.8 GHz Dual-Core processor) and a GPU (NVIDIA GTX1080). The calculation time for the first window is considered separately for the GPU computations since this window also includes the weight initialisation. For each evaluated category, the best performing architecture is highlighted with a bold font.}
%\begin{adjustwidth}{-8pt}{0pt}
\centering
\begin{tabular}{c c c c c}
\toprule
Model & $\#$Param & CPU & GPU 1st window & GPU \\
 &  &  (s/window) & (s) & (ms/window) \\
\hline
PReLU/8 lay./no context & 11,982,464 & 0.222 & 1.432 & 7.70 \\[-0.2ex]
tanh/8 lay./no context & 11,507,328 & \textbf{0.195} & 1.390 & 7.59 \\[-0.2ex]
tanh/8 lay./context & 11,689,984 & 0.236 & \textbf{1.257} & \textbf{7.27} \\[-0.2ex]
\hline
Transmission Line & N/A & 25.16 & N/A & 16918\\[-0.2ex]
\bottomrule
\end{tabular}
\label{tab:timing}
%\end{adjustwidth}
\end{center}
\end{table}

\section*{Discussion}

This paper details how a hybrid, deep-neural-net and analytical approach can be used to develop a real-time model of human cochlear processing (CoNNear), with performance matching that of human cochlear processing. As we demonstrate here for the first time, neither real-time performance nor biophysically realistic responses need be compromised. CoNNear combines both in a single auditory model, laying the groundwork for a new generation of human-like robotic, speech-recognition and machine hearing applications. Prior work has demonstrated the clear benefits of using biophysically realistic cochlear models as front-ends for auditory applications: e.g. for capturing cochlear compression, \cite{saremi_jasa2016, lyon_2017, jepsen_jasa2011}, speech enhancement at negative signal-to-noise ratio's \cite{dbaby_interspeech2015}, realistic sound perception predictions \cite{Verhulst2018b} and for simulating the generation of human auditory brainstem responses \cite{verhulst_jasa2015, verhulst_hearres2018}. Hence, CoNNear can dramatically improve performance in application areas which, to date, have relied on computationally intensive biophysical models. Not only can CoNNear operate on running audio input with a latency below 7.5 ms, the model offers a differentiable solution which can be used in closed-loop systems for auditory feature enhancement or augmented hearing.         

With the rise of neural-network (NN) based methods, computational neuroscience has seen an opportunity to map audio or auditory brain signals directly to sound perception\cite{Kell2018, Akbari2019, Kell2019} and to develop computationally efficient methods to compute large-scale differential-equation-based neuronal networks \cite{amsalem2020}. These developments are transformative, as they can unravel the functional role of hard-to-probe brain areas in perception and yield computationally fast neuromorphic applications. The key to these breakthroughs is the hybrid approach in which knowledge from neuroscience is combined with that of NN-architectures \cite{Richards2019}. While the possibilities of NN approaches are numerous when large amounts of training data are available, this is rarely the case for biological systems and human-extracted data. It therefore remains challenging to develop models of biophysical systems which can generalise to a broad range of unseen conditions or stimuli.

Our work presents a solution to this problem for cochlear processing by constraining the CoNNear architecture and its hyperparameters on the basis of a state-of-the-art TL cochlear model. Our general approach consists of four steps: First, derive an analytical description of the biophysical system on the basis of available experimental data. Second, use the analytical model to generate a training data set consisting of responses to a representative range of relevant stimuli. Third, use this training data set to determine the NN-model architecture and constrain its hyperparameters. Finally, verify the ability of the model to generalise to unseen inputs. Here, we demonstrated that CoNNear predictions generalise to a diverse set of audio stimuli by faithfully predicting key cochlear mechanics features to sounds which were excluded from the training. We note that CoNNear performed best on unseen speech and cochlear mechanics stimuli with levels below 80 dB. We expect that improved generalisability to complex non-speech audio such as environmental sounds and music can be obtained when retraining CoNNear with a more diverse audio set.

Our proposed method is by no means limited to NN-based models of cochlear processing. Indeed, the method can be widely applied to other nonlinear and/or coupled biophysical models of sensory and biophysical systems. Over the years, analytical descriptions of cochlear processing have evolved based on the available experimental data from human and animal cochleae, and they will continue to improve. It is straightforward to train CoNNear to an updated/improved analytical model in step (i), as well as to include different or additional training data in (ii) to further optimise its performance.  
 
\section*{Conclusion}
We present a hybrid method which uniquely combines expert knowledge from the fields of computational auditory neuroscience and machine-learning-based audio processing to develop a CoNNear model of human cochlear processing. CoNNear presents an architecture with differentiable equations and operates in real time ($<$ 7.5 ms delay) at speeds 2000 times faster than state-of-the-art biophysically realistic models. We have high hopes that the CoNNear framework will inspire a new generation of human-like machine hearing, augmented hearing and automatic speech-recognition systems. 

\section*{Methods}

Extended Data Fig.~1 describes the parameters which define the CNN-based CoNNear auto-encoder architecture. The encoder layers transform audio input of size $L$ into a condensed representation of size $L/2^{M} \times k_{M}$, where $k_M$ equals the number of filters in the $M$\textsuperscript{th} CNN layer. We used $128$ filters per layer with a filter length of $64$ samples. The encoder layers use strided convolutions, i.e. the filters were shifted by a time-step of two to halve the temporal dimension after every CNN layer. The decoder contains $M$ deconvolution layers to re-obtain the original temporal dimension of the audio input ($L$). The first three decoder layers used 128 filters per layer and the final decoder layer had ($N_{\text{CF}}$) filters, equalling the number of cochlear sections simulated in the reference cochlear TL model.

\subsection*{Training CoNNear}

CoNNear was trained using TL-model simulations of $N_{\text{CF}}$ BM displacement waveforms in response to audio input windows of $L=2048$ samples. For the context architecture (Fig.~\ref{fig:aecnn}b), the previous and following $L_l$=$L_r$=$256$ input samples were also available to CoNNear. This architecture included a final cropping layer to remove the context after the last CNN decoder and yield an output size $L \times N_{\text{CF}}$. Note that the CoNNear model output units are BM displacement $y_{BM}$ in [$\mu$m], whereas the TL-model outputs are in [m]. This scaling was necessary to enforce a training procedure with sufficiently high digital numbers. For training purposes, and visual comparison between the TL and CoNNear outputs, the $y_{BM}$ values of the TL model were multiplied by a factor of $10^{6}$ in all figures and analyses.

The TIMIT speech corpus \cite{timit} was used for training and contains phonetically balanced sentences with sufficient acoustic diversity. To generate the training data, $2310$ sentences from the TIMIT corpus were used. They were upsampled from $16$ kHz to $100$ kHz to solve the TL-model accurately \cite{altoe_jasa2014} and the root-mean square (RMS) energy of every utterance was adjusted to $70$ dB sound pressure level (SPL). $550$ sentences of the TIMIT dataset were omitted from the training and considered as the test set. BM displacements were simulated for $1000$ cochlear sections with centre frequencies between $25$ Hz and $20$ kHz using a nonlinear time-domain TL model of the cochlea \cite{verhulst_hearres2018}. From the TL-model output representation (i.e., 1000 $y_{BM}$ waveforms sampled at $20$ kHz), outputs from $201$ CFs between $100$ Hz and $12$ kHz, spaced according to the Greenwood map \cite{Greenwood1990}, were chosen to train CoNNear. Above 12 kHz, human hearing sensitivity becomes very poor \cite{ISO2003}, motivating our choice for the upper limit of considered CFs. CoNNear model parameters were optimised to minimise the mean absolute error (dubbed L1 loss) between the predicted model outputs and the reference TL model outputs. A learning rate of $0.0001$ was used with Adam optimiser \cite{adam_optimizer} and the entire framework was developed using the Keras machine learning library \cite{keras} with a Tensorflow \cite{tensorflow} back-end.

\subsection*{Cochlear mechanics evaluation metrics} \label{sec:evaluation}
The evaluation stimuli were sampled at 20 kHz and had a duration of 102.4 ms ($2048$ samples) and 128 ms ($2560$ samples) for the CoNNear and CoNNear-context model, respectively. Stimulus levels were adjusted using the reference pressure of $p_{0}=2\cdot 10^{-5}$Pa. Only when evaluating how CoNNear generalised to longer duration continuous stimuli (Extended Data Fig.~4), or when investigating its real-time capabilities, did we deviate from this procedure. The following sections describe the four cochlear mechanics evaluation metrics we considered to evaluate the CoNNear predictions. Together, these metrics form a comprehensive description of cochlear processing.

\subsubsection*{Cochlear filter tuning} 
A common approach to characterise auditory or cochlear filters is by means of the equivalent-rectangular bandwidth (ERB) or Q\textsubscript{ERB}. The ERB describes the bandwidth of a rectangular filter which passes the same total power than the filter shape estimated from behavioural or cochlear tuning curve experiments \cite{moore1983suggested}, and presents a standardised way to characterise the tuning of the asymmetric auditory/cochlear filter shapes. The ERB is commonly used to describe the frequency and level-dependence of human cochlear filtering \cite{glasberg1990derivation, shera_2002revised, shera2010otoacoustic}, and Q\textsubscript{ERB} to describe level-dependent cochlear filter characteristics from BM impulse response data \cite{verhulst_jasa2012, raufer2016otoacoustic}. We calculated Q\textsubscript{ERB} as:
\begin{equation}
Q_{\text{ERB}} = \frac{\text{CF}}{\text{ERB}}.
\end{equation}
The ERB was determined from the power spectrum of a simulated BM time-domain response to an acoustic click stimulus using the following steps \cite{raufer2016otoacoustic}: (i) compute the Fast Fourier Transform of the BM displacement at the considered CF, (ii) compute the area underneath the power spectrum, and (iii) divide the area by the CF. The frequency- and level-dependence of CoNNear predicted cochlear filters were compared against TL-model predictions and experimental Q\textsubscript{ERB} values reported for humans \cite{shera_2002revised}. 

The acoustic stimuli were condensation clicks of 100-$\mu$s duration and were scaled to the desired peak-equivalent sound pressure level (dB peSPL), to yield a peak-to-peak click amplitude that matched that of a pure-tone with the same dB SPL level ($L$): 
\begin{equation}
\label{eqn:click}
\text{click}(t) = 2 \sqrt{2} \cdot p\textsubscript{0} \cdot 10^{L/20} \cdot x(t) \; \text{with} \; x(t)=
\begin{cases}
1 & \text{for} \: t \leq 100 \, \mu s \\
0 & \text{for} \: t > 100 \, \mu s \\
\end{cases}
\end{equation}

\subsubsection*{Cochlear Excitation Patterns}
Cochlear excitation patterns can be constructed from the RMS energy of the BM displacement or velocity at each measured CF in response to tonal stimuli of different levels. Cochlear excitation patterns show a characteristic half-octave basal-ward shift of their maxima as stimulus level increases \cite{ren2002longitudinal}. Cochlear excitation patterns also reflect the level-dependent nonlinear compressive growth of BM-responses when stimulating the cochlea with a pure-tone of the same frequency as the CF of the cochlear measurement site \cite{robles_2001}. Cochlear pure-tone transfer-functions and excitation patterns have in several studies been used to describe the level-dependence and tuning properties of cochlear mechanics \cite{ narayan1998frequency, robles_2001, ren2002longitudinal}. We calculated excitation patterns for all $201$ simulated BM displacement waveforms in response to pure tones of 0.5, 1 and 2 kHz frequencies and levels between 10 and 90 dB SPL using:  
    \begin{equation}
\text{tone}(t) = p\textsubscript{0} \cdot \sqrt{2} \cdot 10^{L/20} \cdot \sin(2\pi f_{\text{tone}} t),
    \end{equation}
where $t$ corresponds to a time vector of $2048$ samples, $L$ to the desired RMS level in dB SPL, and $f_{tone}$ to the stimulus frequencies. The pure-tones were multiplied with a Hanning-shaped $10$-ms on- and offset ramp to ensure a gradual onset. 

\subsubsection*{Cochlear Dispersion}
Click stimuli can also be used to characterise the cochlear dispersion properties, as their short duration allows for an easy separation of the cochlear response from the evoking stimulus. At the same time, the broad frequency spectrum of the click excites a large portion of the BM. Cochlear dispersion stems from the longitudinal-coupling and tuning properties of BM mechanics \cite{ramamoorthy2010biophysical} and is observed through later click response onsets for BM responses associated with more apical CFs. In humans, the cochlear dispersion delay mounts up to 10-12 ms for stimulus frequencies associated with apical processing \cite{dau2000auditory}. Here, we use clicks of various sound intensities to evaluate whether CoNNear produced cochlear dispersion and BM click responses in line with predictions from the TL-model. 

\subsubsection*{Distortion-product otoacoustic emissions (DPOAEs)}
DPOAEs can be recorded in the ear-canal using a sensitive microphone and are evoked by two pure-tones with frequencies $f_1$ and $f_2$ and SPLs of $L_1$ and $L_2$, respectively. For pure tones with frequency ratios between $1.1$ and $1.3$ \cite{neely_jasa2009}, local nonlinear cochlear interactions generate distortion products, which can be seen in the ear-canal recordings as frequency components which were not originally present in the stimulus. Their strength and shape depends on the properties of the compressive cochlear nonlinearity associated with the electro-mechanical properties of cochlear outer-hair-cells \cite{robles1991two}, and the most prominent DPOAEs appear at frequencies of $2f_2 - f_1$ and $2f_1 - f_2$. Even though CoNNear was not designed or trained to simulate DPs, they form an excellent evaluation metric, as realistically simulated DPOAE properties would demonstrate that CoNNear was able to capture even the epiphenomena associated with cochlear processing.    
As a proxy measure for ear-canal recorded DPOAEs, we considered the BM displacement at the highest simulated CF which, in the real ear, would drive the middle-ear and eardrum to yield the ear-canal pressure waveform in an OAE recording. We compared simulated DPs extracted from the fast Fourier transform of the BM displacement response to simultaneously presented pure tones of $f_1$ and $f_2 = 1.2 f_1$ with levels according to the commonly adopted experimental scissors paradigm: $L_1 = 0.4 L_2 + 39$  \cite{kummer_scissors}. We considered $f_1$ frequencies between 1 and 6 kHz and $L_2$ levels between 10 and 100 dB SPL.

\section*{Additional Information}
DB: conceptualisation, methodology, software, validation, formal analysis, investigation, data curation, writing: original draft, visualisation; AVDB: software, validation, investigation, writing: visualisation; SV: conceptualisation, methodology, resources, writing: original draft, review \& editing, supervision, project administration, funding acquisition.

\section*{Acknowledgments}
This work was supported by the European Research Council (ERC) under the Horizon 2020 Research and Innovation Programme (grant agreement No 678120 RobSpear). The authors would like to thank Christopher and Sarita Shera for their help with the final edits.

\section*{Competing interests}
A patent application (PCTEP2020065893) was filed by UGent on the basis of the research presented in this manuscript. Inventors on the application are Sarah Verhulst, Deepak Baby, Fotios Drakopoulos and Arthur Van Den Broucke.

\section*{Data availability}
The source code of the TL-model used for training is available via {\tt 10.5281/zenodo.3717431} or \\ {\tt github/HearingTechnology/Verhulstetal2018Model}, the TIMIT speech corpus used for training can be found online \cite{timit}. Most figures in this paper can be reproduced using the CoNNear model repository.

\section*{Code availability}
The code for the trained CoNNear model, including instructions of how to execute it is available from github.com/HearingTechnology/CoNNear\_cochlea or 10.5281/zenodo.4056552. A non-commercial, academic UGent license applies.

\nolinenumbers

%\bibliography{refs} 
%%%%%%%%%%%%%%%%%%%%%%%%%%%%%%
%%%%% BIBLIOGRAPHY %%%%%%%%%%%

%%%%%%%%%%%%%%%%%%%%%%%%%%%%%%
%%%%%%%%%%%%%%%%%%%%%%%%%%%%%%

\clearpage
\section*{Extended Data}

\begin{figure*}[h!t]
{\begin{center}
\includegraphics[scale=0.9]{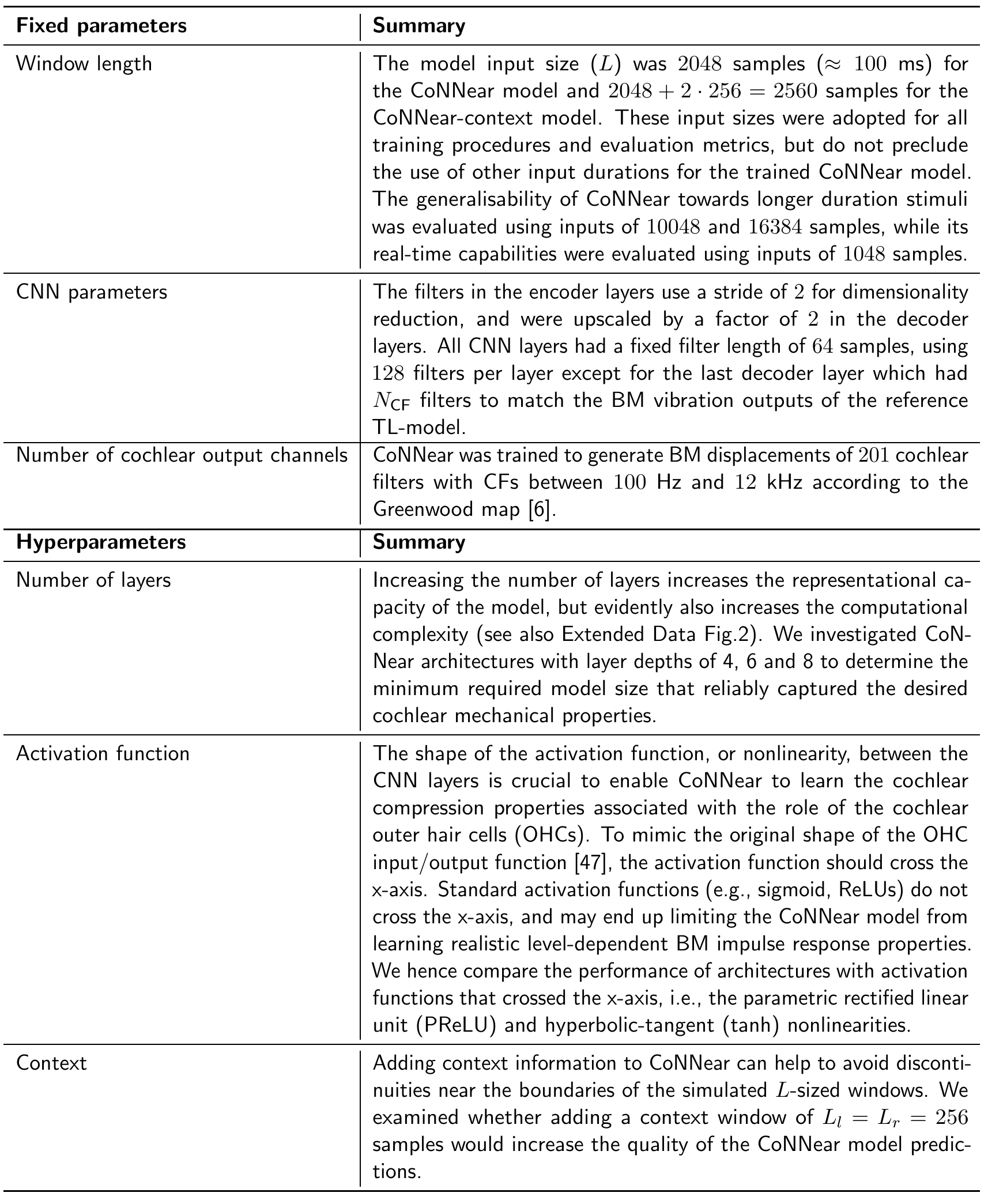}
\end{center}
}
%\vspace{-10pt}
\caption*{\textbf{Extended Data Fig.~1: Overview of the CoNNear architecture parameters.}}
\label{fig:supfig1}
\end{figure*}

\clearpage

\begin{figure*}[h!t]
{\begin{center}
\includegraphics[scale=1]{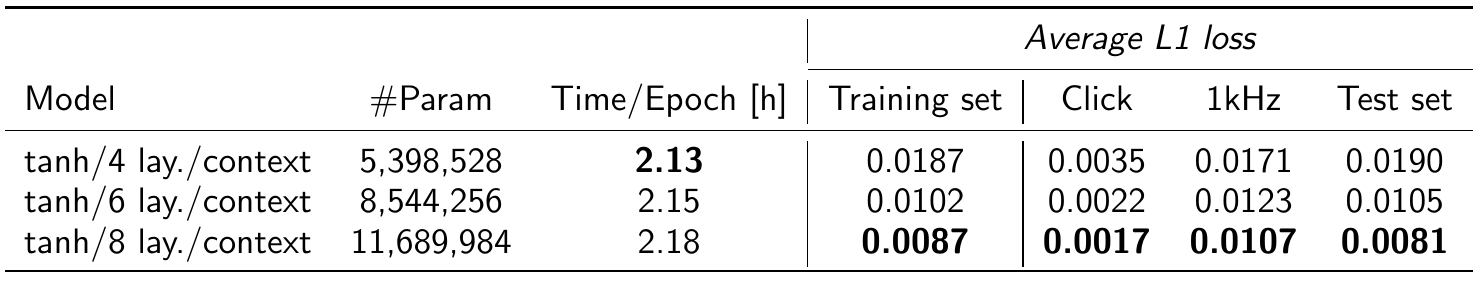}
\end{center}
}
%\vspace{-10pt}
\caption*{\textbf{Extended Data Fig.~2: CNN layer depth comparison.} The first column details the CoNNear architecture. The next columns describe the total number of required model parameters, the required training time per epoch of $2310$ TIMIT training sentences and average L1 loss across all windows of the TIMIT training set. Average L1 losses were also computed for BM displacement predictions to a number of unseen acoustic stimuli (click and 1-kHz pure tones) with levels between 0 and 90 dB SPL. Lastly, average L1 loss was also computed for the 550 sentences of the TIMIT test set. For each evaluated category, the best performing architecture is highlighted in bold font.}
\label{fig:supfig3}
\end{figure*}

%\begin{table}[h!]
%\begin{center}
%\caption*{\textbf{Extended Data Fig. 2: CNN layer depth comparison.} The first column details the CoNNear architecture. The next columns describe the total number of required model parameters, the required training time per epoch of $2310$ TIMIT training sentences and average L1 loss across all windows of the TIMIT training set. Average L1 losses were also computed for BM displacement predictions to a number of unseen acoustic stimuli (click and 1-kHz pure tones) with levels between 0 and 90 dB SPL. Lastly, average L1 loss was also computed for the 550 sentences of the TIMIT test set. For each evaluated category, the best performing architecture is highlighted in bold font.% For each evaluated category, (-) and (+) depict the best and worst performing architecture, respectively.
%}
%\centering
%%\begin{adjustwidth}{-8pt}{0pt}
%\begin{tabular}{l c c |c|c c c}
%\toprule
%& & &  \multicolumn{4}{c}{\textit{Average L1 loss}}\\
%\cmidrule{4-7}
%Model & $\#$Param & Time/Epoch [h] & Training set & Click & 1kHz & Test set \\ 
%\midrule
%tanh/4 lay./context & 5,398,528 & \textbf{2.13}  & 0.0187  & 0.0035 & 0.0171 & 0.0190\\[-0.2ex]
%tanh/6 lay./context & 8,544,256 & 2.15 &  0.0102 & 0.0022 & 0.0123 & 0.0105\\[-0.2ex]
%tanh/8 lay./context & 11,689,984 & 2.18 &  \textbf{0.0087} & \textbf{0.0017} & \textbf{0.0107} & \textbf{0.0081}\\[-0.2ex]
%\bottomrule
%\end{tabular}
%\label{tab:ld}
%%\end{adjustwidth}
%\end{center}
%\end{table}

\clearpage

\begin{figure*}[h!t]
{\begin{center}
\includegraphics[scale=1]{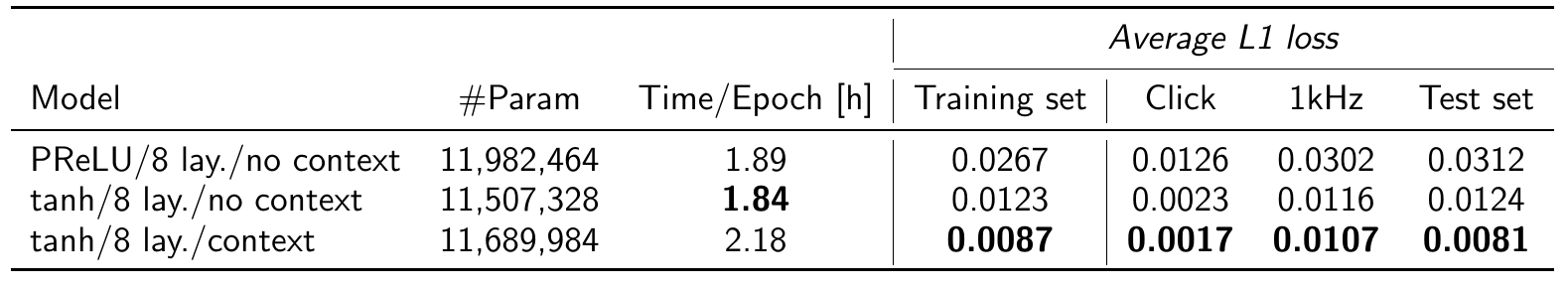}
\end{center}
}
%\vspace{-10pt}
\caption*{\textbf{Extended Data Fig.~3: Activation function comparison.} The first column details the CoNNear architecture. The next columns describe the total number of required model parameters, the required training time per epoch of $2310$ TIMIT training sentences and average L1 loss across all windows of the TIMIT training set. Average L1 losses were also computed for BM displacement predictions to a number of unseen acoustic stimuli (click and 1-kHz pure tones) with levels between 0 and 90 dB SPL. Lastly, average L1 loss was also computed for the 550 sentences of the TIMIT test set. For each evaluated category, the best performing architecture is highlighted in bold font. }
\label{fig:supfig2}
\end{figure*}

%\begin{table}[h!]
%\begin{center}
%\caption*{\textbf{Extended Data Fig. 3: Activation function comparison.} The first column details the CoNNear architecture. The next columns describe the total number of required model parameters, the required training time per epoch of $2310$ TIMIT training sentences and average L1 loss across all windows of the TIMIT training set. Average L1 losses were also computed for BM displacement predictions to a number of unseen acoustic stimuli (click and 1-kHz pure tones) with levels between 0 and 90 dB SPL. Lastly, average L1 loss was also computed for the 550 sentences of the TIMIT test set. For each evaluated category, the best performing architecture is highlighted in bold font. %(-) and (+) depict the best and worst performing architecture, respectively.
%}
%\begin{adjustwidth}{-8pt}{0pt}
%\centering
%\begin{tabular}{l c c |c|c c c}
%\toprule
%& & & \multicolumn{4}{c}{\textit{Average L1 loss}}\\
%\cmidrule{4-7}
%Model & $\#$Param & Time/Epoch [h] & Training set & Click & 1kHz & Test set \\ 
%\midrule
%PReLU/8 lay./no context & 11,982,464 & 1.89  & 0.0267  & 0.0126 & 0.0302 & 0.0312\\[-0.2ex]
%tanh/8 lay./no context & 11,507,328 & \textbf{1.84} &  0.0123 & 0.0023 & 0.0116 & 0.0124\\[-0.2ex]
%tanh/8 lay./context & 11,689,984 & 2.18 &  \textbf{0.0087} & \textbf{0.0017} & \textbf{0.0107} & \textbf{0.0081}\\[-0.2ex]
%\bottomrule 
%\end{tabular}
%\label{tab:nl}
%%\end{adjustwidth}
%\end{center}
%\end{table}

\clearpage
\begin{figure*}[h!]
{\begin{center}
\includegraphics[scale=1]{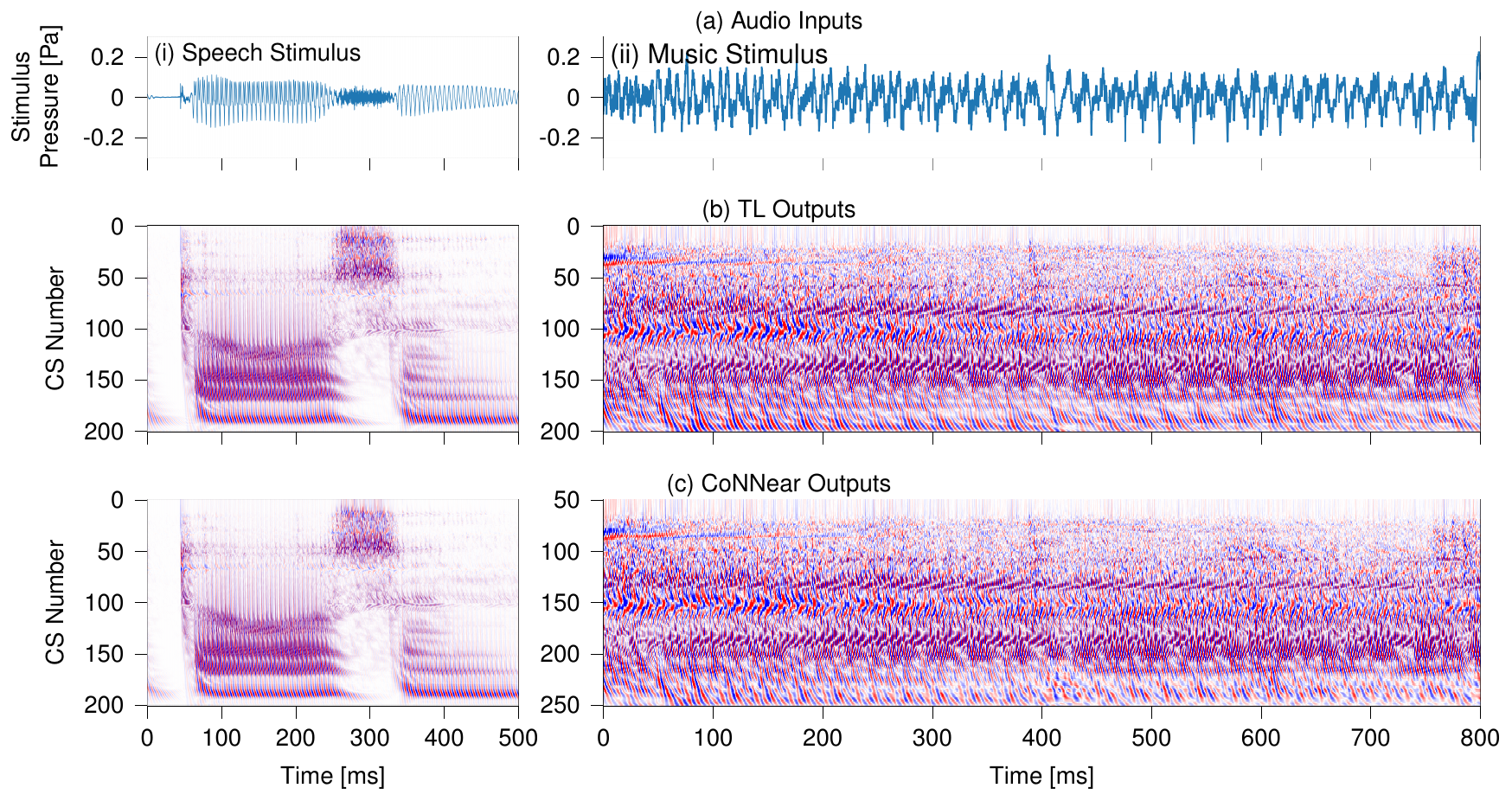}
\end{center}
}
\vspace{-10pt}
\caption*{\textbf{Extended Data Fig.~4: Simulated BM displacements for a 10048-sample speech stimulus and a 16384-sample music stimulus.} The stimulus waveform is depicted in panel (a) and panels (b)-(c) depict instantaneous BM displacement intensities (darker colours = higher intensities) of the simulated TL-model (b) and CoNNear (c) outputs. The N\textsubscript{CF}=201 considered output channels are labelled per channel number: channel 1 corresponds to a CF of 12 kHz and channel 201 to a CF of 100 Hz. The same colour scale was used for both simulations and ranged between -0.5 $\mu$m (blue) and 0.5 $\mu$m (red). The left panels show simulations to a speech stimulus from the Dutch matrix test \cite{houben2014development} and the right panels shows simulations to a music fragment (Radiohead - No Surprises).}
\label{fig:speechlong}
\end{figure*}
%Here we should make a new page for each additional figure, and its (extended)legend

\clearpage
\begin{figure*}[h!]
{\centering
\includegraphics[scale=1.]{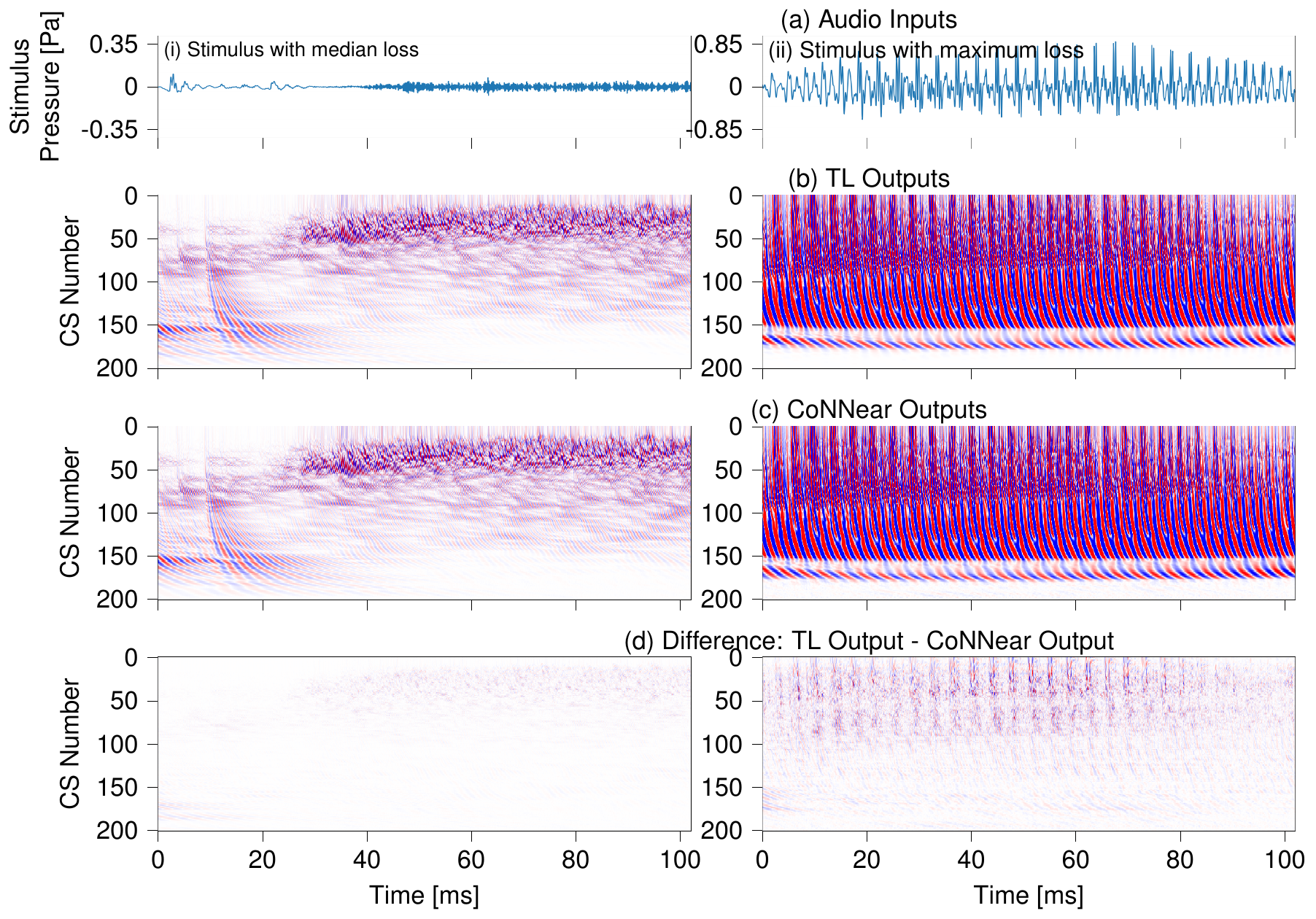}
}
\caption*{\textbf{Extended Data Fig.~5: Comparing TL and CoNNear model predictions at the median and maximum L1 prediction error.} The figure visually compares BM displacement intensities of the BM (b) and CoNNear (c) model to audio fragments which resulted in the median and maximum L1 errors of 0.008 and 0.038 simulated for the TIMIT test set (Fig.~\ref{fig:histogram_and_examples}). The N\textsubscript{CF}=201 considered output channels are labelled per channel number: channel 1 corresponds to a CF of 12 kHz and channel 201 to a CF of 100 Hz. The same colour scale was used for both simulations and ranged between -0.5 $\mu$m (blue) and 0.5 $\mu$m (red).}
\label{fig:median_max_losses}
\end{figure*}

\clearpage
\begin{figure*}[h!]
{\centering
\includegraphics[scale=1.]{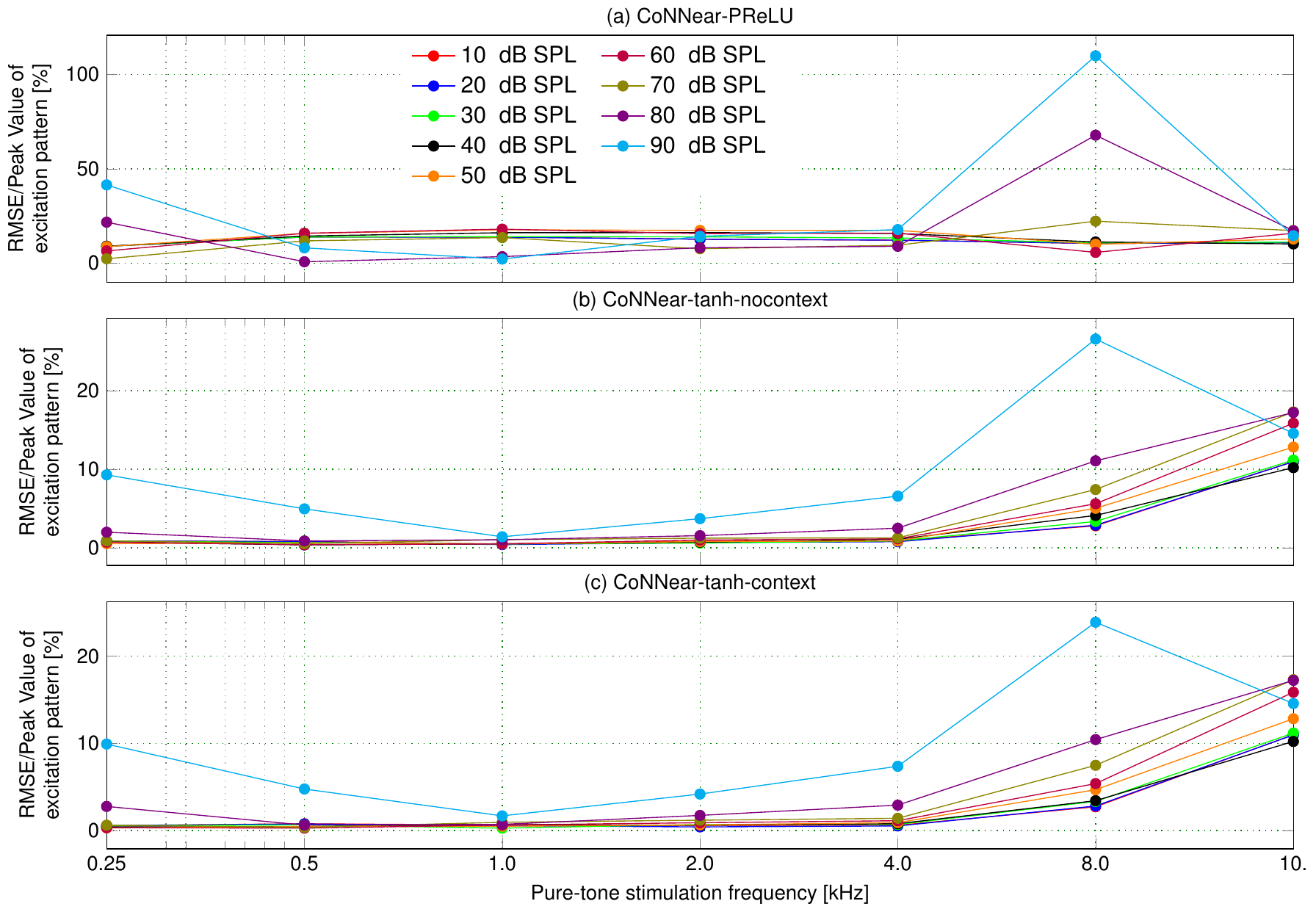}
}
\caption*{\textbf{Extended Data Fig.~6: Root mean-square error (RMSE) between simulated excitation patterns of the TL and CoNNear models reported as fraction of the TL excitation pattern maximum (cf. Fig.~\ref{fig:expat}).} Using the PReLU activation function (a) leads to an overall high RMSE as this architecture failed to learn the level-dependent cochlear compression characteristics and filter shapes. The models using the tanh nonlinearity (b),(c) did learn to capture the level-dependent properties of cochlear excitation patterns, and performed with errors below 5\% for the frequency ranges and stimulus levels captured by the speech training data (for CFs below 5 kHz, and stimulation levels below 90 dB SPL) The RMSE increased above 5\% for all architectures when evaluating its performance on 8- and 10-kHz excitation patterns. This decreased performance results from the limited frequency content of the TIMIT training material.}
\label{fig:rmse_expat}
\end{figure*}

\end{document}